\documentclass[12pt]
  {article}
\usepackage{verbatim,graphicx,amsmath,calc,epsf,epsfig,subfigure}

\input{babarsym.tex}
\newcommand{\deltaE} {\ensuremath{\Delta \rm{E}}\xspace}
\newcommand{\Brec}   {\ensuremath{\B_{\rm{rec}}}\xspace}
\newcommand{\Bopp}   {\ensuremath{\B_{\rm{opp}}}\xspace}
\newcommand{\Bcp}    {\ensuremath{\B_{\rm{CP}}}\xspace}
\newcommand{\Btag}   {\ensuremath{\B_{\rm{tag}}}\xspace}

\def\Dstarp   {\ensuremath{D^{*+}}\xspace}
\def\Dstarm   {\ensuremath{D^{*-}}\xspace}

\def\DorDstarm {\ensuremath{D^{(*)-}}\xspace}
\def\DorDstarzb  {\ensuremath{\Dbar^{(*)0}}\xspace}

\def\pipi     {\ensuremath{\pi\pi}}

\def\hh     {\ensuremath{h^+h^{\prime -}}}

\def\fpm {\ensuremath{f_{\pm}(\deltat)}}
\def\ilam {\ensuremath{{\cal I}m\lambda}}
\def\alam {\ensuremath{\left|\lambda\right|}}

\def\spipi {\ensuremath{S_{\pi\pi}}}
\def\cpipi {\ensuremath{C_{\pi\pi}}}
\def\de {\ensuremath{\Delta E}}

\def\Bflav {\ensuremath{B_{\rm flav}}}

\newcommand\hips{\mbox{${\rm ps}^{-1}$}}

\newcommand{\BABARPubYear}    {01}

\newcommand{\BABARProcNumber} {86}
\newcommand{\SLACPubNumber} {9075}

\long\def\inst#1{\par\nobreak\kern 4pt\nobreak
    {\it #1}\par\vskip 10pt plus 3pt minus 3pt}

\begin{document}
{\pagestyle{empty}

\begin{flushright}
SLAC-PUB-\SLACPubNumber \\
\babar-PROC-\BABARPubYear/\BABARProcNumber \\
December, 2001 \\
\end{flushright}

\par\vskip 4cm

\begin{center}
\Large \bf Measurements of \CP Violation, Mixing and Lifetimes of $B$ Mesons
      with the \babar\ Detector
\end{center}
\bigskip

\begin{center}
\large 
S\"oren Prell\\
University of California at San Diego\\
Department of Physics\\
9500 Gilman Drive, La Jolla, CA 92093\\
(for the \lbabar\ Collaboration)
\end{center}
\bigskip \bigskip

\begin{center}
\large \bf Abstract
\end{center}
We report the observation of \CP\ violation in the $B^0$ meson
system. Using a novel technique for time-dependent measurements, we
measure a non-zero value for the \CP -violating amplitude \stwob at the
4.1 $\sigma$ level. We also report on precision measurements of the \Bp
and \Bz lifetimes and the \BzBzb mixing frequency \deltamd obtained with
the same technique, and on a  first measurement of the time-dependent
\CP-violating amplitude in $B^0\to \pip\pim$ decays. 

\vfill
\begin{center}
Contributed to the Proceedings of the\\ 9$^{th}$ International 
Symposium on Heavy Flavor Physics, \\
9/10/2001---9/13/2001, Caltech, Pasadena
\end{center}

\vspace{1.0cm}
\begin{center}
{\em Stanford Linear Accelerator Center, Stanford University, 
Stanford, CA 94309} \\ \vspace{0.1cm}\hrule\vspace{0.1cm}
Work supported in part by Department of Energy contract DE-AC03-76SF00515.
\end{center}

\clearpage
}

\section{Introduction}
\CP\ violation has been a central concern of particle physics since its 
discovery in 1964 in the decays of \KL\ mesons~\cite{KLCP}. 
An elegant explanation of the origin of \CP\ violation was proposed by
Kobayashi and Maskawa, as a complex phase in the three-generation 
CKM quark-mixing matrix~\cite{CKM}. 
In this picture, measurements of \CP-violating
asymmetries in the time distributions of  
\Bz decays to charmonium final states are expected to be large and
provide a direct test of the Standard Model of electroweak
interactions~\cite{BCP}. 

We present measurements of time-dependent \CP-asymmetries in samples
of fully reconstructed $B$ decays to charmonium-containing \CP\
eigenstates ($b\to\ccbar s$) and to the $\pi^+\pi^-$ final state. The
data for these studies were recorded at the $\FourS$
resonance by the \babar\ detector at the PEP-II asymmetric-energy \epem
collider at the Stanford Linear Accelerator Center.

When the \FourS\ decays, the $P$-wave \BzBzb\ state evolves coherently
until one of the mesons decays. In one of four time-order and flavor
configurations,  if the tagging meson $B_{\rm tag}$ decays first, and as
a \Bz, the other meson must be a \Bzb\ at that same time $t_{\rm
  tag}$. It then evolves independently, and can decay into a \CP
eigenstate $B_{CP}$ at a later time $t_{CP}$. The time between the two
decays $\deltat=t_{CP}-t_{\rm tag}$ is a signed quantity made measurable
by producing the $\FourS$ with a boost $\beta\gamma=0.56$ along the
collision ($z$) axis, with nominal energies of 9.0 and 3.1\gev\ for the
electron and positron beams. The measured distance
$\deltaz\approx\beta\gamma c\deltat$ between the two decay vertices
provides a good estimate of the corresponding time interval \deltat;
the average value of $|\deltaz|$ is  $\beta\gamma c\tau_{\Bz}\approx
250\mum$.   

We examine each of the events in the  $B_{\CP}$ sample for  
evidence that the other neutral $B$ meson decayed as a \Bz or a \Bzb
(flavor tag). The distribution ${\rm f}_+({\rm f}_-)$ of the decay rate  
when the tagging meson is a $\Bz (\Bzb)$ is given by
\begin{eqnarray}
\fpm = \frac{e^{-\left|\deltat\right|/\tau}}{4\tau} [1
\pm  S\sin(\deltamd\deltat) 
\mp  C\cos(\deltamd\deltat)],
\label{eq:timedist}
\end{eqnarray}
where $\deltat$ is the time between the two \B decays, 
$\tau$ is the \Bz lifetime~\cite{PDG2000}, \deltamd is the \BzBzb mixing
frequency~\cite{PDG2000}, and the lifetime difference between neutral
\B\ mass eigenstates is assumed to be negligible.  
The sine term in Eq.~\ref{eq:timedist} is due to interference between 
direct decay and decay after mixing, and the cosine term is due to direct \CP\ violation.
The \CP-violating parameters $S$ and $C$ are defined in terms
of a complex parameter $\lambda$ that
depends on both \BzBzb mixing and on the amplitudes describing \Bzb and
\Bz decay to a common final state $f$~\cite{lambda}:
\begin{equation}
S = \frac{2\,\ilam}{1+\alam^2}\quad{\rm and}\quad C = \frac{1-\alam^2}{1+\alam^2}.
\label{SandCdef}
\end{equation}
A difference between the \Bz and \Bzb \deltat distributions or
a \deltat asymmetry for either flavor tag is evidence for \CP violation.

In the Standard Model $\lambda=\eta_f e^{-2i\beta}$ for
charmonium-containing $b\to\ccbar s$ decays, $\eta_f$ is the \CP eigenvalue of
the state $f$ and
$\beta = \arg \left[\, -V_{\rm cd}^{}V_{\rm cb}^* / V_{\rm td}^{}V_{\rm tb}^*\, \right]$
is an angle of the Unitarity Triangle of the three-generation CKM matrix~\cite{CKM}.
Thus, the time-dependent \CP-violating asymmetry is
\begin{eqnarray}
A_{\CP}(\deltat) \equiv  \frac{ {\rm f}_+(\deltat)  -  {\rm f}_-(\deltat) }
{ {\rm f}_+(\deltat) + {\rm f}_-(\deltat) } 
= -\eta_f \stwob \sin{ (\Delta m_{B^0} \, \deltat )} . 
\label{eq:asymmetry}
\end{eqnarray}

The analogous time-dependent $\CP$-violating asymmetry in the decay
$\Bz\to\pip\pim$ arises from interference between mixing and decay
amplitudes, and interference between the $b\to uW^-$ (tree) and $b\to
dg$ (penguin) decay amplitudes. If the decay proceeds purely through the
tree process, the complex parameter $\lambda$ is directly related to CKM
matrix elements and $\alam = 1$ and $\ilam = \stwoa$, where $\alpha
= \arg\left[-V_{td}V_{tb}^*/V_{ud}V_{ub}^*\right]$.   

However, recent theoretical estimates suggest that the contribution from the
gluonic penguin amplitude can be significant~\cite{Beneke01a} 
leading to $\alam\ne 1$ and $\ilam = \alam\sin{2\alpha_{\rm eff}}$, 
where $\alpha_{\rm eff}$ depends on the magnitudes and strong phases of
the tree and penguin amplitudes.  

We also present precise measurements of the \Bz-\Bzb\ mixing frequency
\deltamd\ and the neutral and charged $B$ lifetimes. These measurements
use the same vertexing algorithm and \deltat calculation as the measurements
of the \CP-asymmetries. In addition, for the \deltamd\ measurement, the
same flavor tagging algorithm as for the \CP analyses is used. These
measurements are amongst the most precise available and provide a good
validation of the novel technique to study time-dependent $B$ decays.

In all analyses the values of the parameters under study ($B$ lifetimes,
\deltamd , \stwob , $\spipi$ and $\cpipi$ ) were hidden to eliminate possible
experimenter's bias until event selection, \deltat~reconstruction
method, and fitting procedures were finalized and systematic errors were
determined.  

\section{The \babar\ Detector}
A detailed description of the \babar\ detector can be found in
Ref.~\cite{ref:babar}. Charged particles are detected and their momenta
measured by a combination of a silicon vertex tracker (SVT) consisting
of five double-sided layers and a central drift chamber (DCH), in a
1.5-T solenoidal field. The average vertex resolution in the $z$
direction is 70\mum\ for a fully reconstructed $B$ meson. We identify
leptons and hadrons with measurements from all detector systems,
including the energy loss (\dedx) in the DCH and SVT. Electrons
and photons are identified by a CsI electromagnetic calorimeter
(EMC). Muons are identified in the instrumented flux return (IFR).
A Cherenkov ring imaging detector (DIRC) covering the central region,
together with the \dedx\ information, provides $K$-$\pi$ separation of at
least three standard deviations for $B$ decay products with momentum
greater than 250\mevc in the laboratory.

\section{Measurement of $B$ Lifetimes and \deltamd }
The measurements of the charged and neutral $B$ lifetimes and the
\Bz-\Bzb\ mixing frequency \deltamd\ are based on a sample of
approximately 23 million \BB\ pairs. 

\subsection{Exclusive $B$ Reconstruction}
Samples of \Bz\ and \Bu\ mesons \Brec are reconstructed 
in the modes $\Bz \ra \DorDstarm \pip$, $\DorDstarm \rho^+$, $\DorDstarm
a_1^+$, $\jpsi  \Kstarz$ and $\Bu \ra \DorDstarzb \pip$, $\jpsi K^+$,
$\psitwos K^+ $.
Charged and neutral \Dstarb\ candidates are formed by combining a \Dzb
with a \pim\ or \piz . \Dzb\ candidates are reconstructed in the decay
channels $\Kp\pim$, $\Kp\pim\piz$, $\Kp\pip\pim\pim$ and $\KS \pip\pim$
and \Dm\ candidates in the decay channels $\Kp \pim \pim$ and $\KS\pim$.  
We reconstruct \jpsi and \psitwos in the decays to $\epem$ and
$\mu^+\mu^-$ and the \psitwos decay to \jpsi \pip \pim . 

\begin{figure}[htbp]
\label{fig:hadronic}
\includegraphics[height=0.35\textwidth]{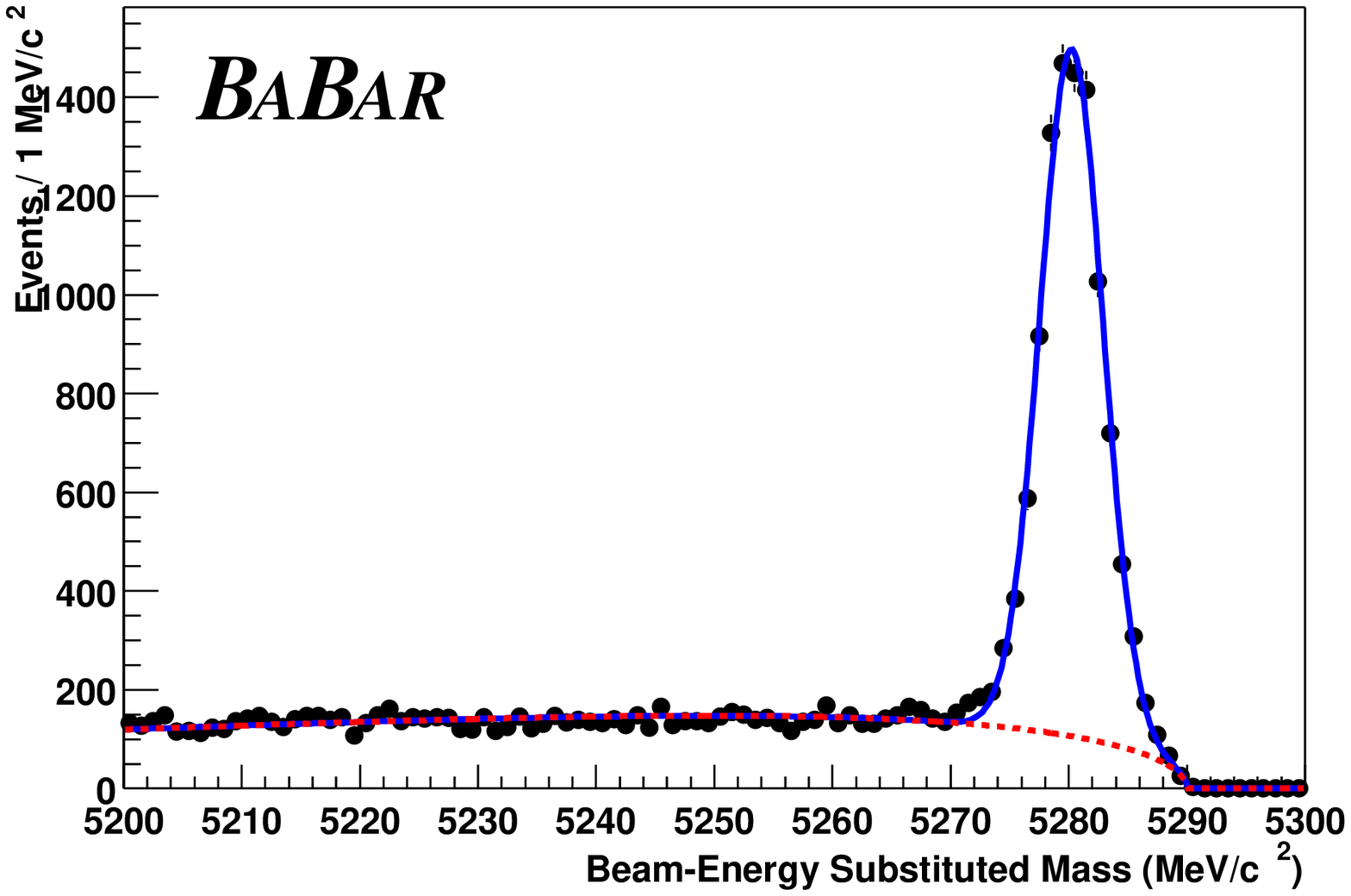}
\includegraphics[height=0.35\textwidth]{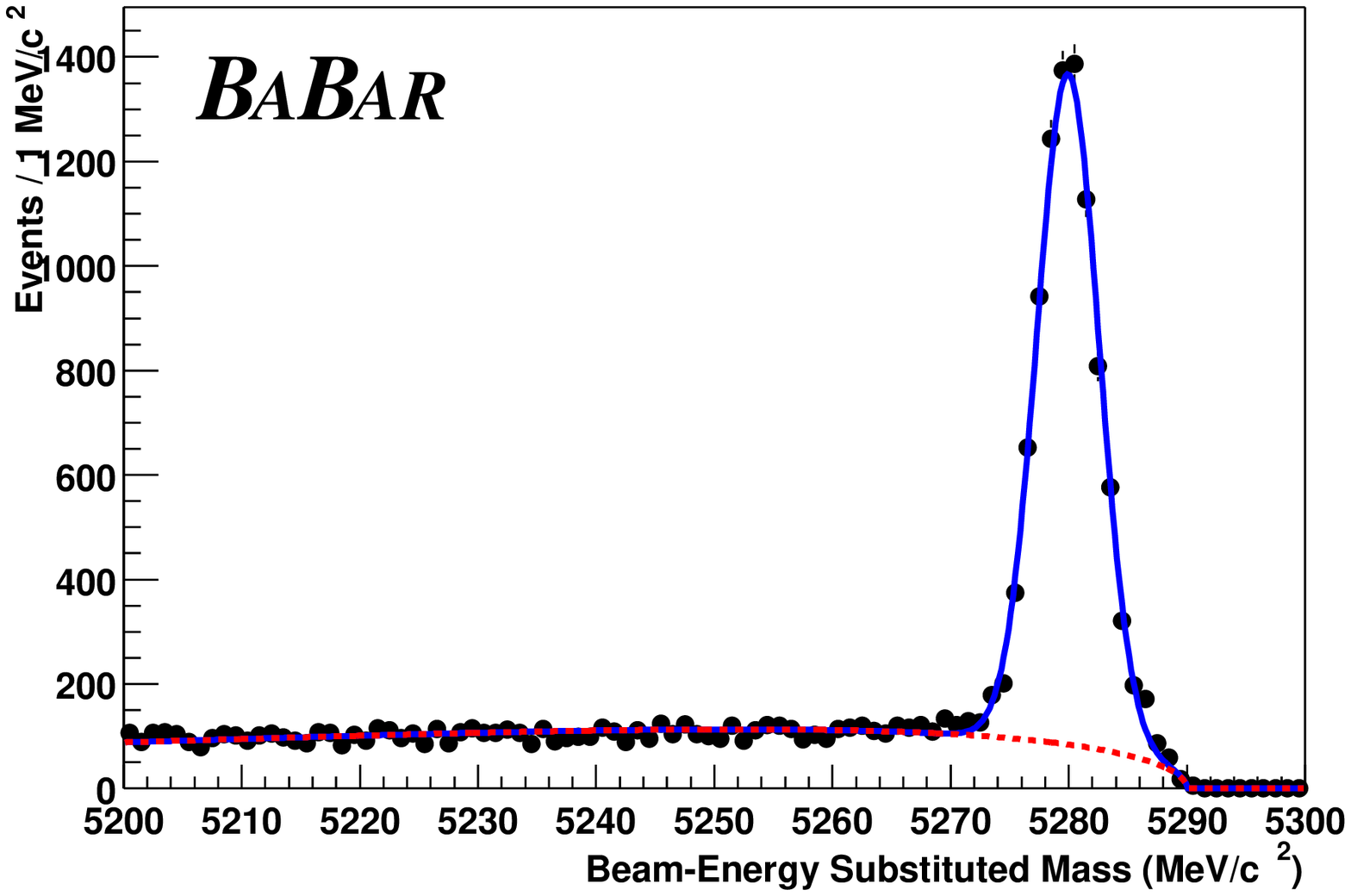}
\caption{Beam energy substituted mass distribution for selected $B^0$
  (left) and $B^+$ (right) candidates. In 30 fb$^{-1}$, we reconstruct 9400
  \Bz and 8500 \Bp  signal events. Average signal purities for
  $\mes>5.27\gevcc$, are 83~\% and  85~\% for \Bz and \Bp , respectively. }
\end{figure}

Continuum $e^+e^-\to q\overline q$ background is suppressed by
requirements on the normalized second Fox-Wolfram moment~\cite{ref:foxw}
for the event and on the angle between the thrust axes of \Brec\ and of
the other $B=$\Bopp in the event. $B$ candidates are identified by 
the difference \deltaE between the reconstructed $B$ energy 
and the beam energy~$\sqrt{s}/2$ in the \FourS\ frame, and the
beam-energy substituted mass
\mes\ calculated from $\sqrt{s}/2$ and the reconstructed $B$
momentum. We require $\mes > 5.2$\gevcc 
and $|\deltaE| < 3 \sigma_{\deltaE}$. The distributions of \mes\ for
selected $B$ candidates in 30 fb$^{-1}$ is shown in Fig.~\ref{fig:hadronic}.

\subsection{\deltat Measurement}
\label{sec:decaytime}
The decay time difference, \deltat, between \B\ decays is determined
from the measured separation  $\Delta z = z_{rec}-z_{opp}$ along the $z$
axis between the \Brec and the \Bopp vertices. This $\Delta z$ is
converted into \deltat using the \FourS\ boost and correcting on an
event-by-event basis for the direction of the \B\ mesons.  The
resolution of the \deltat\ measurement is dominated by the $z$
resolution of the \Bopp decay vertex. This vertex uses all tracks in the
event except those incorporated in $B_{rec}$. An additional constraint
is provided by a calculated \Bopp production point and
three-momentum, determined from the three-momentum of the $B_{rec}$
candidate, its decay vertex, and the average position of the interaction
point and the \FourS\ boost. Reconstructed $\KS$ or $\Lambda$
candidates are used as input to the fit in place of their daughters in
order to reduce bias due to long-lived particles. Tracks with a large
contribution to the $\chi^2$ are iteratively removed from the fit, 
until all remaining tracks have a reasonable fit probability. Candidates 
with $|\Delta z|<3.0$\mm\ and $\sigma_{\Delta z}<400$\mum\ are retained.
For the measurement of the $B$ lifetimes, we require that at least
two tracks are included in the \Bopp vertex .

Two different parameterizations are used to model the decay-time
difference resolution functions. In the measurements of \deltamd ,
\stwob , and \stwoa , the resolution function is approximated by a
sum of three Gaussian distributions with different means $\delta_{k}$
and widths $\sigma_k$,   
\begin{equation}
{\cal R}(\delta_{t},\sigma_{\deltat}| \hat {a} ) =  \sum_{k=1}^{3} 
{ \frac{f_k}{\sigma_k\sqrt{2\pi}} {\rm e}^{ 
-(\delta_{t}-\delta_{k} )^2/2{\sigma_k}^2} },
\end{equation}
where $\delta_{\rm t}$ is the difference between the
measured and true $\deltat$ values. 
For the core and tail Gaussians, the widths
$\sigma_{1,2}=S_{1,2}\times\sigma_{\deltat}$ are the scaled event-by-event 
measurement error, $\sigma_{\deltat}$, derived from the vertex fits. 
The third Gaussian, with a fixed width of $\sigma_3=8$\ps, accounts for 
less than 1\% of {\it outlier} events with
incorrectly reconstructed vertices. 
The three Gaussian resolution function is not suited for the measurement
of the $B$ lifetimes because strong correlations lead to increased
statistical errors. Studies with Monte Carlo simulation and data
show that the sum of a zero-mean Gaussian distribution and its
convolution with an exponential provides a good trade-off between
statistical and systematic uncertainties: \par 
\small{
\vspace*{-6mm}
\begin{eqnarray}
\label{eq:resol}
{\cal R} (\delta_{t} , \sigma_{\deltat} | \hat{a}= \{ h,s,\kappa \}) = 
h    \frac{1}{\sqrt{2\pi}  s \sigma_{\deltat}}   \exp\left(
-\frac{\delta_{t}^2}{2s^2 \sigma_{\deltat}^2}
\right) \ \ \ \ \ \ \ \ \ \ \ \ \ \\
 + \int_{-\infty}^{0} \frac{1-h}{\kappa\sigma_{\deltat}}  \exp\left(
\frac{\delta_{t}'}{\kappa\sigma_{\deltat}} \right)   
\frac{1}{\sqrt{2\pi}  s \sigma_{\deltat}}   \exp\left(
-\frac{(\delta_{t} - \delta_{t}')^2}{2s^2 \sigma_{\deltat}^2}\right)
{\rm d}(\delta_{t}')\; .\nonumber 
\end{eqnarray}
}
\normalsize
\hspace*{-5pt}
The parameters $\hat{a}$ are  
the fraction $h$ in the core Gaussian component, 
a scale factor $s$ for the per-event errors $\sigma_{\deltat}$, and the 
factor~$\kappa$ in the effective time constant~$\kappa\sigma_{\deltat}$
of the exponential which accounts for charm decays. $\Delta t$ {\it
  outlier} events are modeled the same way as in the three Gaussian
resolution function. 
The resolution function parameters are assumed to be the same for all
\Bz and \Bp decay modes. This assumption is confirmed by Monte Carlo
simulation studies. The resolution functions differ only slightly
between \Bz and \Bu mesons due to different mixtures of $D^-$ and \Dzb\
mesons in the  \Bopp decays and we use a single set of resolution
function parameters for both \Bz\ and~\Bu\ in the lifetime fits.

\subsection{$B$ Lifetime Results}
We extract the \Bp and \Bz lifetimes from an unbinned maximum likelihood
fit to the \deltat distributions of the selected $B$ candidates. The
probability for an event to be signal is estimated from $\mes$~fits  
(Fig.~\ref{fig:hadronic}) and the \mes\ value of the \Brec\ candidate.
In the likelihood, the probability density for the signal events is
given by  
\begin{equation}
\label{eq:Phi}
{\cal G}(\deltat,\sigma_{\deltat} | \tau,\hat{a}) = \int_{-\infty}^{+\infty}
e^{-|\deltat |/\tau}/(2\tau)
\, \calR \it (\deltat-\deltat', \sigma_{\deltat} |\hat{a}) \; {\rm d}(\deltat'),
\end{equation}
and the background \deltat\ distribution for each \B\ species
is empirically modeled by the sum of a prompt component and a lifetime
component convolved with the same resolution function, but with a
separate set of parameters. 
The likelihood fit involves 17 free parameters in addition to the \Bz
and the \Bp lifetimes: 12 to describe the background \deltat
distributions and 5 for the signal resolution function.
The charged $B$ lifetime $\tau_{\Bu }$ is replaced with $\tau_{\Bu } =
r \cdot \tau_{\Bz }$ to estimate the statistical error on the
ratio~$r= \tau_{\Bu}/\tau_{\Bz}$.  

We determine the \Bz and \Bu meson lifetimes and their ratio to be:
\begin{eqnarray}
 \tau_{\Bz} &=& 1.546 \pm 0.032\mbox{ (stat)} \pm 0.022\mbox{ (syst)} \mbox{ ps,}\nonumber \\
 \tau_{\Bu} &=& 1.673 \pm 0.032\mbox{ (stat)} \pm 0.023\mbox{ (syst)} \mbox{ ps, and}\nonumber \\
 \tau_{\Bu}/\tau_{\Bz} &=& 1.082 \pm 0.026\mbox{ (stat)} \pm 0.012\mbox{ (syst)}.\nonumber 
\end{eqnarray}

\begin{figure}[htbp]
\label{fig:dt-life-n-mix} 
\begin{center}
\includegraphics[height=.31\textheight]{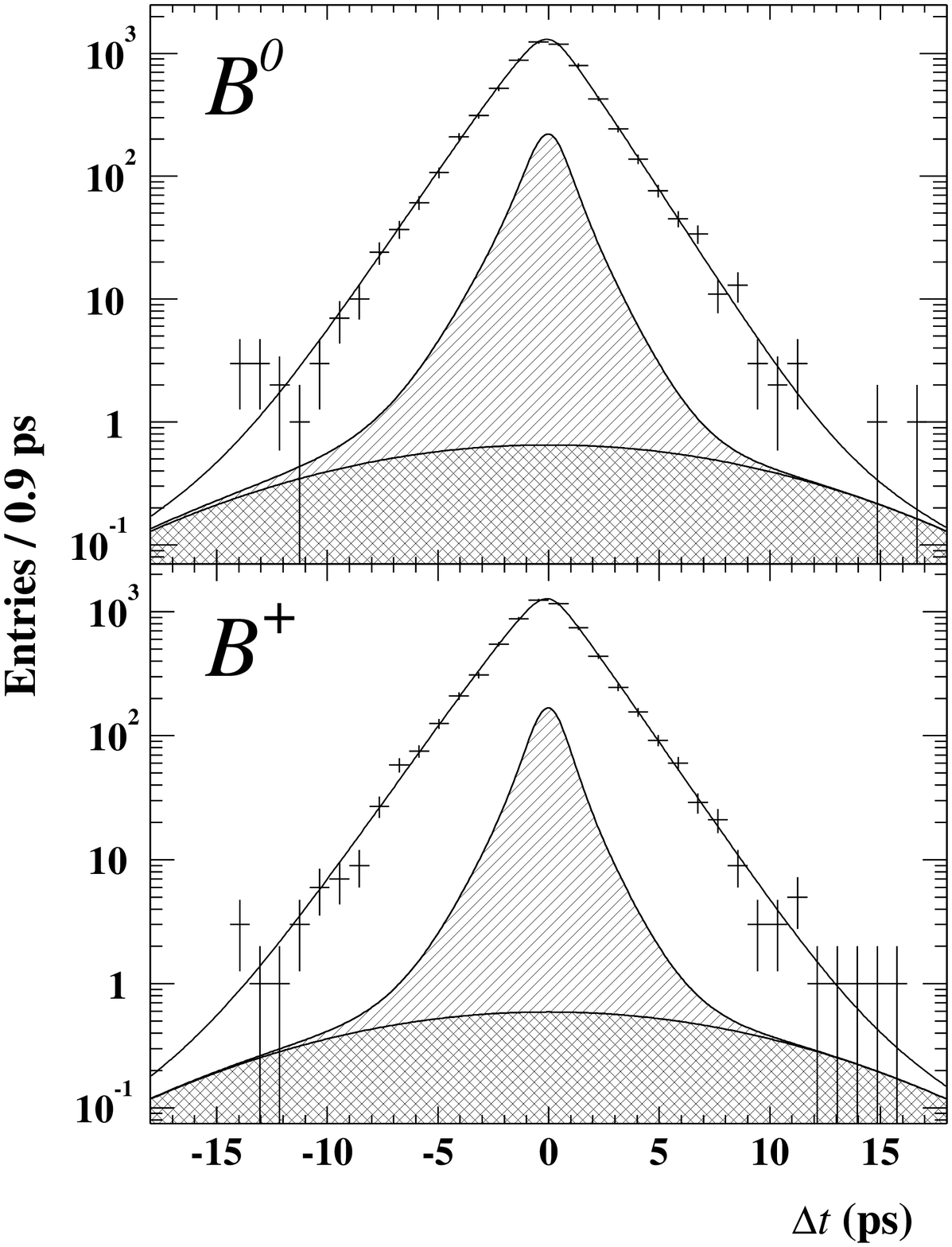}
\includegraphics[height=.33\textheight]{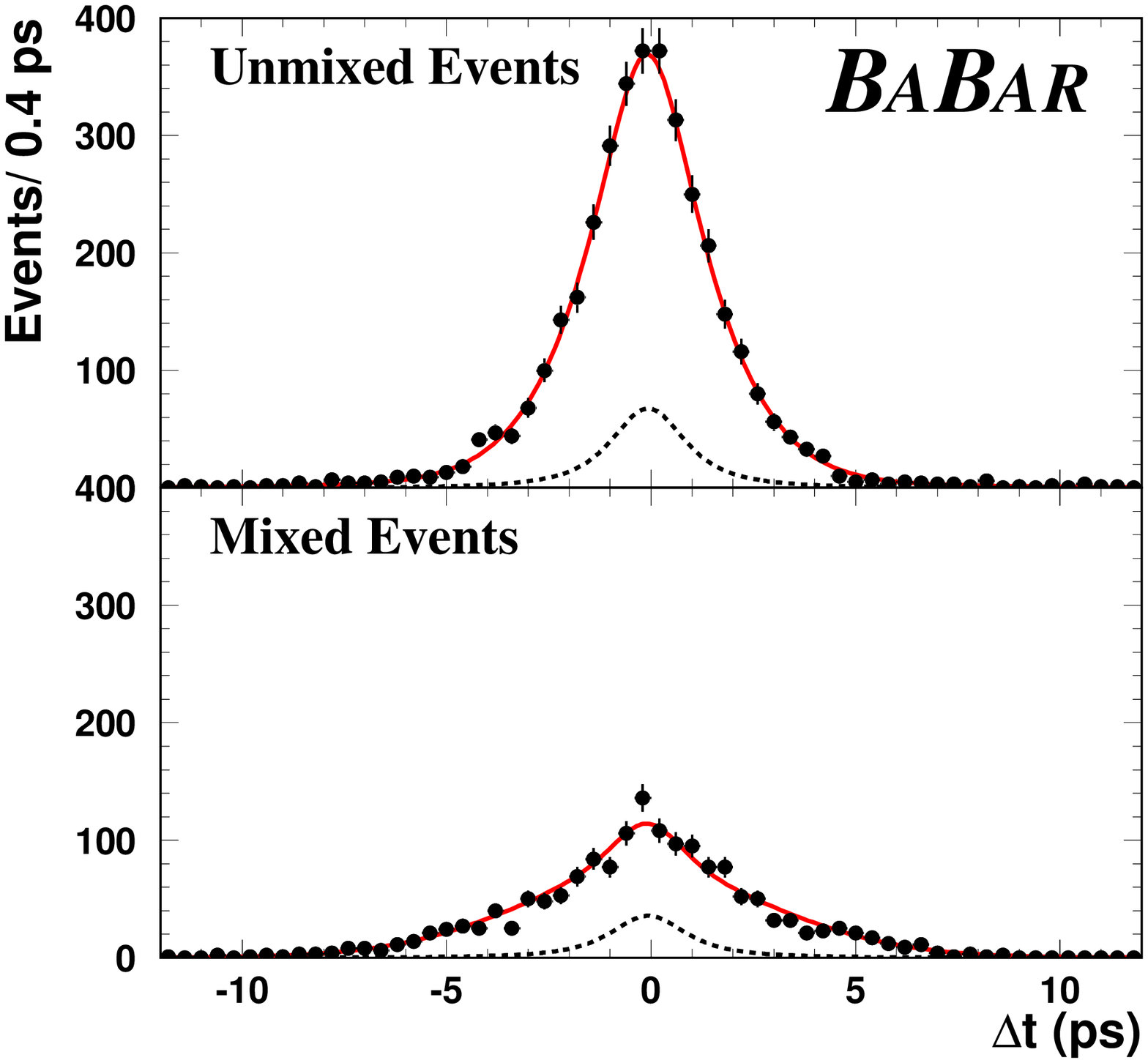}
\end{center}
\caption{Left: \deltat\ distribution for the \Bz\ (top) and \Bu\ (bottom) 
events within $2 \sigma$ of the \B mass with 
superimposed fit results. The single-hatched areas are the background
contributions and the cross-hatched areas represent the \deltat
outliers. The probability of obtaining a lower likelihood is 7.3\%.
Right: \deltat\ distributions in data for the mixed and unmixed
    events ($\mes(B_{rec})>5.27$\gevcc), with overlaid the projection of the
    likelihood fit (solid) and the background contributions (dashed).}
\end{figure}

These are the most precise measurements to date~\cite{babar-blife} and
are consistent with the world averages~\cite{PDG2000}. The resolution
function parameters are consistent with those found in a Monte Carlo
simulation that includes detector alignment effects. 
Figure~\ref{fig:dt-life-n-mix} shows the results of the likelihood fit 
superimposed on the \deltat~distributions for \Bz\ and \Bu\ events. 
With the current data sample these measurements are still statistically 
limited. The dominant systematic errors arise from uncertainties in the 
description of the combinatorial background and of events with large
\deltat\ values, the use of a common time resolution function for $B^0$
and $B^+$ and from limited Monte Carlo statistics.

\subsection{$B$ Flavor Tagging}
The measurements of \deltamd, \stwob and \stwoa require knowledge of the
\Bopp = \Btag flavor. We use the same tagging algorithm in the three
analyses to determine the \Btag flavor from the charges of its decay
products.   

The charge of energetic electrons and muons from semileptonic $B$
decays, kaons, soft pions from \Dstar decays, and  high momentum charged
particles is correlated with the flavor of the decaying $b$ quark.  
Each event is assigned to one of four hierarchical,
mutually exclusive 
tagging categories or has no flavor tag.
A lepton tag requires an electron (muon) candidate with a 
center-of-mass momentum $p_{\rm cm} >1.0\ (1.1)\gevc$. This efficiently selects
primary leptons and
reduces contamination due to oppositely-charged leptons from charm decays.
Events meeting these criteria are assigned to the {\tt Lepton} 
category unless the lepton charge and the net charge of all kaon candidates indicate opposite tags. 
Events without a lepton tag but with a non-zero net kaon charge are assigned to the 
{\tt Kaon} category. 
All remaining
events are passed to a neural network algorithm 
whose main inputs are the momentum and charge 
of the track with the highest center-of-mass 
momentum, and the outputs of secondary networks,
trained with Monte Carlo samples to identify primary leptons, kaons, 
and soft pions. 
Based on the output of the neural network algorithm, events are tagged as \Bz or \Bzb and 
assigned to the {\tt NT1} (more certain tags) or {\tt NT2} (less certain tags) category, or 
not tagged at all. The tagging power of the {\tt NT1} and {\tt NT2}
categories arises primarily from soft pions and from recovering
unidentified isolated primary electrons and muons.  
The yields, efficiencies, purities and mistag rates $w$ for each tagging
category are listed in Table.~\ref{tab:HadronicBYield}.

\begin{table}[!htb]
\caption{
Event yields for the different tagging categories obtained from fits to
the \mes\ distributions.  The purity is quoted for $\mes >5.270 \mevcc$
and average mistag fractions $\mistag_i$ 
extracted for each tagging category $i$ 
from the likelihood fit to the time distribution for the  
fully-reconstructed flavor eigenstate sample.}
\begin{center}
\begin{tabular}{lcccc} \hline
Category     & Tagged       & Efficiency (\%) & Purity (\%) & \mistag \\ \hline
{\tt Lepton} & $ 754\pm 28$ & $11.3\pm 0.4$    & $97.1\pm 0.6$ & $0.085\pm 0.018$ \\
{\tt Kaon}   & $2317\pm 54$ & $34.8\pm 0.6$    & $85.2\pm 0.8$ & $0.167\pm 0.014$ \\
{\tt NT1}    & $ 556\pm 26$ & $ 8.3\pm 0.3$    & $88.7\pm 1.5$ & $0.195\pm 0.026$ \\ 
{\tt NT2}    & $ 910\pm 36$ & $13.7\pm 0.4$    & $83.0\pm 1.3$ & $0.326\pm 0.024$ \\ \hline
Total        & $4538\pm 75$ & $68.1\pm 0.9$    & $86.7\pm 0.5$  \\ \hline
\end{tabular}
\end{center}
\label{tab:HadronicBYield}
\end{table}

\subsection{\BzBzb  Mixing Result}
The value of \deltamd\ is extracted from the tagged flavor-eigenstate
\Bz sample \Bflav\ with a simultaneous unbinned likelihood fit to the
\deltat\ distributions of both unmixed ($\BzBzb$) and mixed ($\Bz\Bz$
and $\Bzb\Bzb$) events. The PDFs for the unmixed $(+)$ and mixed $(-)$
signal events for the $i^{th}$ tagging category are given by
\begin{equation}
{\cal H}_{\pm}(\deltat| \deltamd, \mistag_i, \hat{a}_i)  = \frac{{\rm e}^{ -\left| \deltat \right|/\tau}}{4\tau}
\left[ 1 \pm (1-2\mistag_i)\cos{ \deltamd \deltat } \right] \otimes {\cal R}(\delta_t|\hat {a}_i),
\end{equation}
Some resolution function parameters are allowed to differ for each
tagging category to account for shifts due to inclusion of charm decay
products in the tag vertex. The PDFs are extended to include background
terms, different for each tagging category. 
The probability that a \Bz\ candidate is a signal event is determined
from a fit to the observed \mes\ distribution for its tagging category.
The \deltat\ distributions of the combinatorial background are described
with a zero lifetime component and a non-oscillatory component with
non-zero lifetime. Separate resolution function parameters are used for
signal and background to minimize correlations.

The likelihood fit involves a total of 34 parameters, including 
\deltamd (1), the mistag rates $\mistag_i$ and mistag differences
$\Delta\mistag _i=\mistag _i(\Bz)-\mistag _i(\Bzb)$ (8),  
\deltat resolution function parameters (9) and background parameters
(16). We display the result of the likelihood fit by using the mixing
asymmetry, 
\begin{equation}
{\cal A}_{mix}(\deltat) =
\frac{N_{unmixed}(\deltat)-N_{mixed}(\deltat)}{N_{unmixed}(\deltat)+N_{mixed}(\deltat)}. 
\label{eq:asym}
\end{equation}

\begin{figure}
\label{fig:mix_asym}
\begin{center}
\includegraphics[height=.35\textheight]{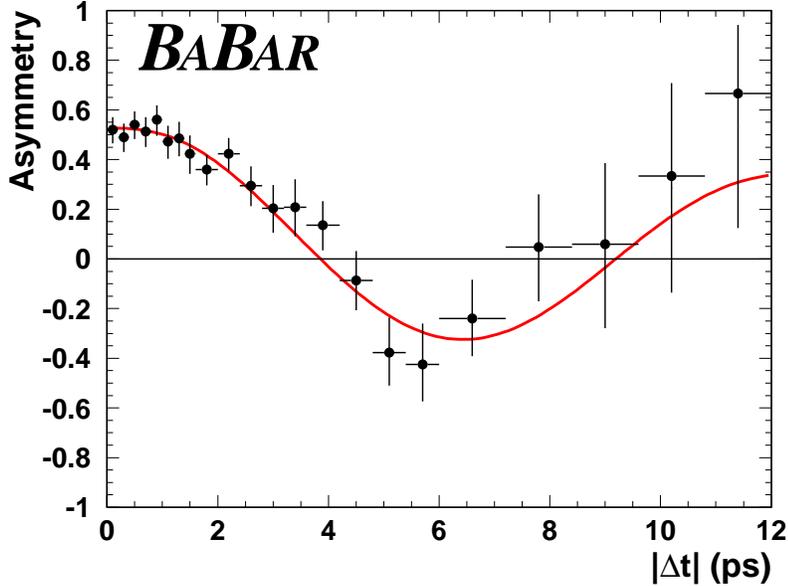}%
\end{center}
\caption{The asymmetry ${\cal A}_{mix} (\deltat)$ between unmixed and
  mixed events as a function of $|\deltat|$ overlaid with the result
  from the likelihood fit. 
} 
\end{figure}

If flavor tagging and \deltat\ determination were perfect, the asymmetry
as a function of \deltat\ would be a cosine with unit amplitude. The
amplitude is diluted by mistag probabilities and the \deltat
resolution. The \deltat\ distributions of mixed and unmixed events, and
their asymmetry, ${\cal A}_{mix}$, 
are shown in Fig.~\ref{fig:dt-life-n-mix} and~\ref{fig:mix_asym}
along with projections of the fit result. The probability to obtain a
smaller likelihood is $28~\%$.    

Systematic uncertainties in the \deltamd measurement arise from various
sources. The conversion of \deltaz\ to \deltat\ introduces an
uncertainty ($\pm 0.007\,\ps^{-1}$) due to the limited knowledge of the
PEP-II boost, the $z$ length scale of \babar\ and the $B_{\rm rec}$
momentum vector in the \FourS\ frame. Systematic uncertainties related
to the resolution function ($\pm 0.005\,\ps^{-1}$), are attributed to
the choice of the parameterization, the description of outliers, and the
capability of the resolution model to deal with various plausible
misalignment scenarios applied to the Monte Carlo simulation. The
parameters of the background \deltat\ distribution are left free in the
likelihood fit, but systematic errors ($\pm 0.005\,\ps^{-1}$), are
introduced by the uncertainty in signal probabilities, parameterization
of the background \deltat distributions and resolution function, and the
small amount of correlated \Bu\ background. Finally, statistical
limitations of Monte Carlo validation tests ($\pm 0.004\,\ps^{-1}$), the
full size of a (negative) correction obtained from Monte Carlo ($\pm
0.009\,\ps^{-1}$), and the variation of the \Bz\ lifetime~\cite{PDG2000}
($\pm 0.006\,\ps^{-1}$) contribute. These contributions added in
quadrature yield a total systematical error of 0.016 \hips .

In conclusion, the \Bz-\Bzb\ mixing frequency \deltamd\ is determined  
with a new time-dependent technique from tagged fully-reconstructed
\Bz decays to be  
\begin{eqnarray}
\deltamd = 0.519 \pm 0.020 ({stat}) \pm 0.016 ({syst})\,\hips.\nonumber
\end{eqnarray}
This preliminary result is one of the single most precise measurements
available, and is consistent with the current world average~\cite{PDG2000}
and a recent \babar\ measurement with a dilepton sample~\cite{dilepton}.
The error on \deltamd\ is still dominated by statistics, leaving
substantial room for further improvement as more data are accumulated. 

\section{Measurement of $\stwob $}
In 32 million \BB , we extract \stwob from  a sample of fully
reconstructed \Bz decays ($\Bcp$) to final states with 
$\eta_f=-1$ ($\jpsi\KS$, $\psitwos\KS$, $\chicone \KS$), 
$\eta_f=+1$ ($\jpsi\KL$) and $\eta_f({\rm effective}) = 0.65\pm0.07$
($\jpsi\Kstarz $  with $\Kstarz\to\KS\piz$)~\cite{babar-s2b-1}.

\begin{figure}[!bht]
\label{fig:asymlike}
\begin{center}
\includegraphics[height=.38\textheight]{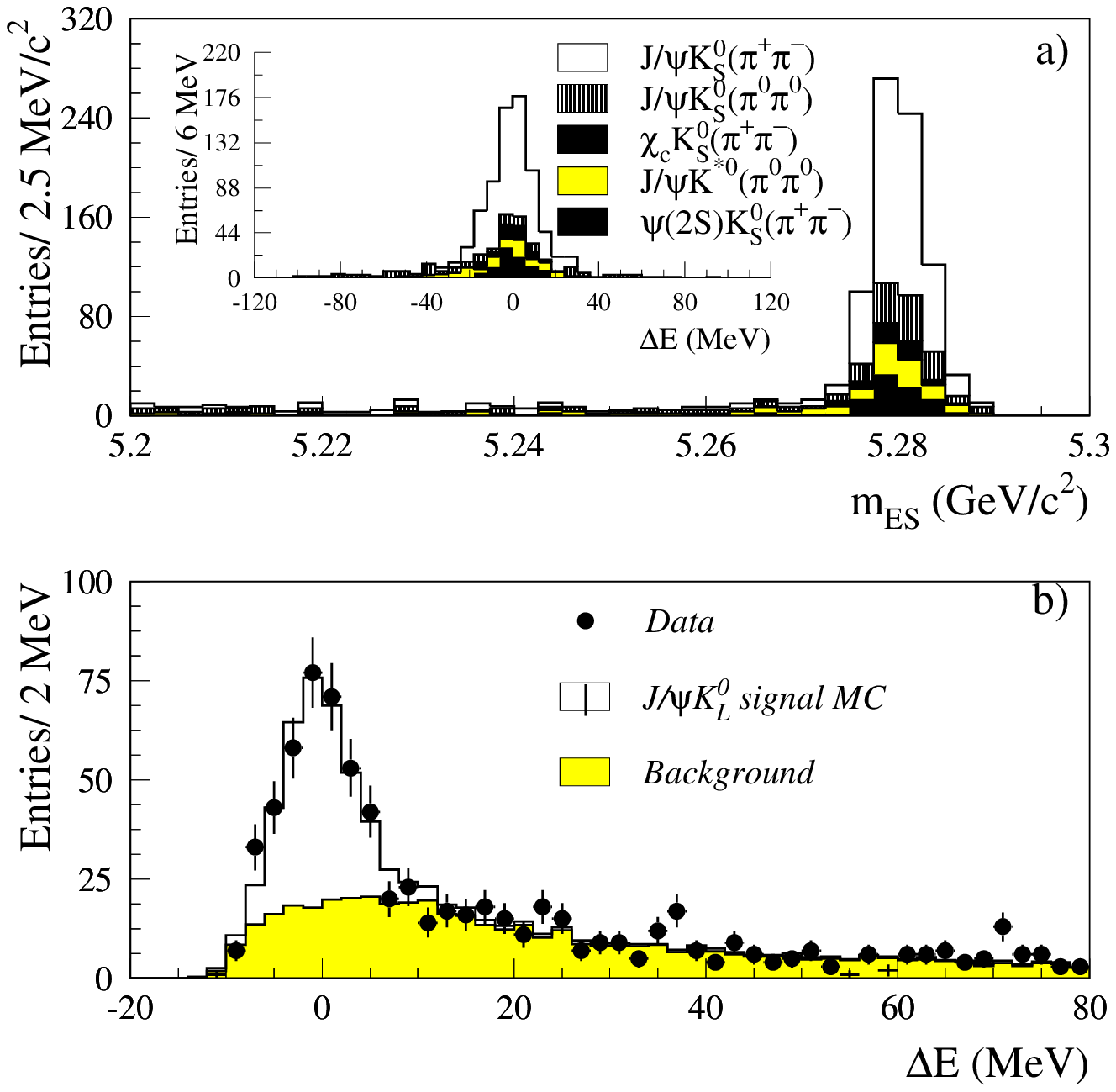}
\includegraphics[height=.38\textheight]{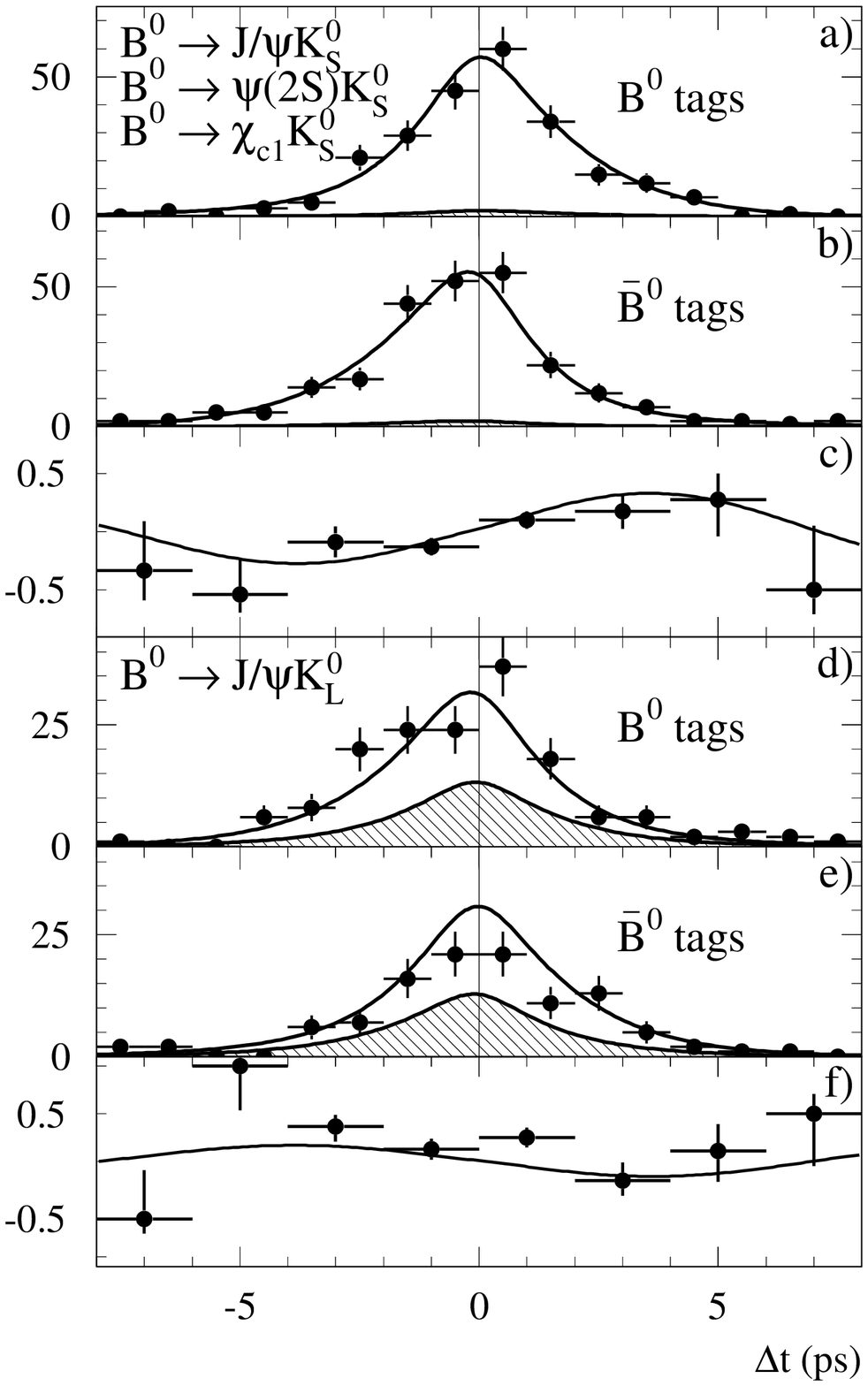}
\end{center}
\caption{Left: a) distribution of \mes\ for $B_{\CP}$
  candidates having a \KS in the final state; 
b) distribution of $\Delta E$ for $\jpsi\KL$ candidates. 
Right: Number of $\eta_f=-1$ candidates ($\jpsi\KS$, $\psitwos\KS$,
  and $\chicone\KS $) in the signal region  a) with a \Bz tag $N_{\Bz} $ 
and b) with a \Bzb tag $N_{\Bzb}$, and c) the asymmetry
$(N_{\Bz}-N_{\Bzb})/(N_{\Bz}+N_{\Bzb})$, as functions of \deltat . The
solid curves represent the result of the combined fit to all selected \CP events;
the shaded regions represent the background contributions.
Figures d)--f) contain the corresponding information for the 
$\eta_f=+1$ mode $(\jpsi\KL)$.}
\end{figure}

The \stwob measurement is made with a simultaneous unbinned likelihood
fit to the \deltat distributions of the tagged $B_{\CP}$ and $B_{\rm
  flav}$ samples. The \deltat\ distribution of the former is given by
Eq.~\ref{eq:timedist}, with $|\lambda|=1$. The  $B_{\rm flav}$ sample
evolves according to the PDF for \Bz flavor oscillations  as described
in the previous section. The amplitudes for the $B_{\CP}$ \CP
-asymmetries and 
for the $B_{\rm flav}$ flavor oscillations are reduced by the same factor
$(1-2\mistag)$ due to wrong tags. Both distributions are convolved with
a common \deltat resolution function and backgrounds are accounted for
by adding terms to the likelihood, incorporated with different
assumptions about their \deltat evolution and convolved with a separate
resolution function. Events are assigned signal and background
probabilities based on the \mes\ (all modes except $\jpsi\KL$) or
$\Delta E$ ($\jpsi\KL$) distributions shown in 
Fig.~\ref{fig:asymlike}. Separate \deltat\ resolution 
functions parameters have been used for the data collected in 1999-2000
and 2001, due to the significant improvement in the SVT alignment. 

A  total of 45 parameters are varied in the likelihood fit, including
\stwob (1), mistag fractions $\mistag$ and differences $\Delta\mistag$
(8), parameters for the signal \deltat resolution (16), and parameters
for background time dependence (9), \deltat resolution (3) and mistag
fractions (8). The determination of the mistag fractions and signal
\deltat resolution function is dominated by the large $B_{\rm flav}$
sample. The largest correlation between \stwob\ and any linear
combination of the other free parameters is only 0.13. We fix
$\tau_{\Bz}$ and \deltamd \cite{PDG2000}. 

Figure~\ref{fig:asymlike} shows the $\deltat$ distributions and
${A}_{\CP}$ as a function of \deltat overlaid with the likelihood fit
result for the $\eta_f = -1$ and $\eta_f = +1$ samples. The probability
of obtaining a lower likelihood is 27\%. The simultaneous fit to all \CP
decay modes and flavor decay modes yields 
\begin{eqnarray}
\stwob=0.59 \pm 0.14\ \stat \pm 0.05\ \syst. \nonumber
\end{eqnarray}
The dominant sources of systematic error are the parameterization of the
\deltat\ resolution function (0.03), due in part to residual
uncertainties in SVT alignment, possible differences in the mistag
fractions between the $B_{\CP}$ and $B_{\rm flav}$ samples (0.03),  
and uncertainties in the level, composition, and \CP\ asymmetry of the
background in the selected \CP events (0.02). The systematic errors from 
uncertainties in $\Delta m_{\Bz}$ and $\tau_{\Bz}$ and from the
parameterization of the background in the $B_{\rm flav}$ sample are
small; an increase of $0.020\,\ps^{-1}$ in the value for $\Delta
m_{\Bz}$ decreases \stwob\ by 0.015.  

\begin{table}[!htb] 
\caption{ 
Number of tagged events, signal purity and observed \CP\ asymmetries in 
the \CP\ samples and control samples.  
Errors are statistical only.}
\label{tab:result} 
\begin{center}
\begin{tabular}{lrrr} \hline
 Sample  & $N_{\rm tag}$ & Purity (\%) & \multicolumn{1}{c}{$\ \ \
 \stwob$}\\
\hline
$\jpsi\KS$,$\psitwos\KS$,$\chicone\KS$   & $480$        & $96$       &  0.56$\pm$0.15   \\ 
$\jpsi \KL$                              & $273$        & $51$       &  0.70$\pm$0.34   \\
$\jpsi\Kstarz ,\Kstarz \to \KS\piz       $& $50$         & $74$       &  0.82$\pm$1.00  \\ 
\hline
 Full \CP\ sample                        & $803$        & $80$       &  0.59$\pm$0.14   \\ 
\hline\hline
$B_{\rm flav}$ non-\CP sample            & $7591$       & $86$       &  0.02$\pm$0.04     \\
\hline 
Charged $B$ non-\CP sample        & $6814$       & $86$       &  0.03$\pm$0.04     \\
\hline
\end{tabular} 
\end{center}
\end{table}

The large sample of reconstructed events allows a number of consistency
checks, including separation of the data by decay mode, tagging
category and $B_{\rm tag}$ flavor. The results of fits to some subsamples
and to the samples of non-\CP decay modes are shown in
Table~\ref{tab:result}. 
For the latter, no statistically significant asymmetry is found. 

If $\vert\lambda\vert$ is allowed to float in the fit to the
$\eta_f=-1$ sample, which has high purity and requires minimal
assumptions on backgrounds, we obtain
$\vert\lambda\vert = 0.93 \pm 0.09\ \stat \pm 0.03\ \syst $. 
The sources of systematic error are the same as in the \stwob analysis. 

The measurement of $\stwob=0.59 \pm 0.14\ \stat \pm 0.05\ \syst $  
establishes \CP violation in the \Bz meson system at the $4.1\sigma$
level. This significance is computed from the sum in quadrature of  
the statistical and additive systematic errors.
The probability of obtaining this value or higher in the absence of \CP violation
is less than 
$3 \times 10^{-5}$. The corresponding probability for the 
$\eta_f=-1$ modes alone is $2 \times 10^{-4}$.

\section{Measurement of $\stwoa $}
We reconstruct neutral $B$ mesons decaying to $\hh$, where $h$ and
$h^{\prime}$ refer to $\pi$ or $K$ in a sample of 33
million \BB. The data set includes $30.4\invfb$ collected on the
\Y4S\ resonance and $3.3\invfb$ collected below the \BB\ threshold used
for continuum background studies.      

\begin{figure}[bht]
\begin{center}
\includegraphics[height=.28\textheight]{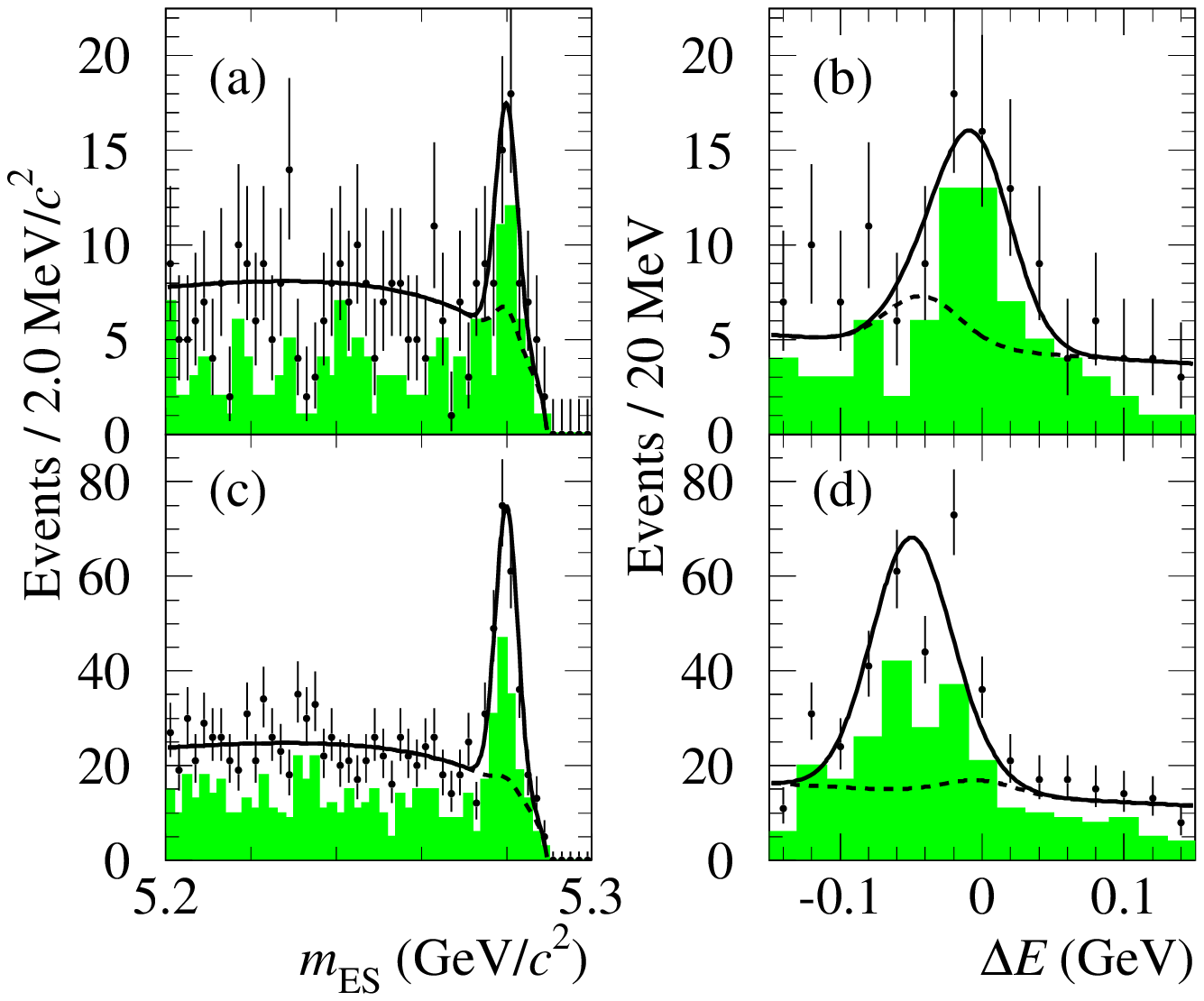}
\includegraphics[height=.28\textheight]{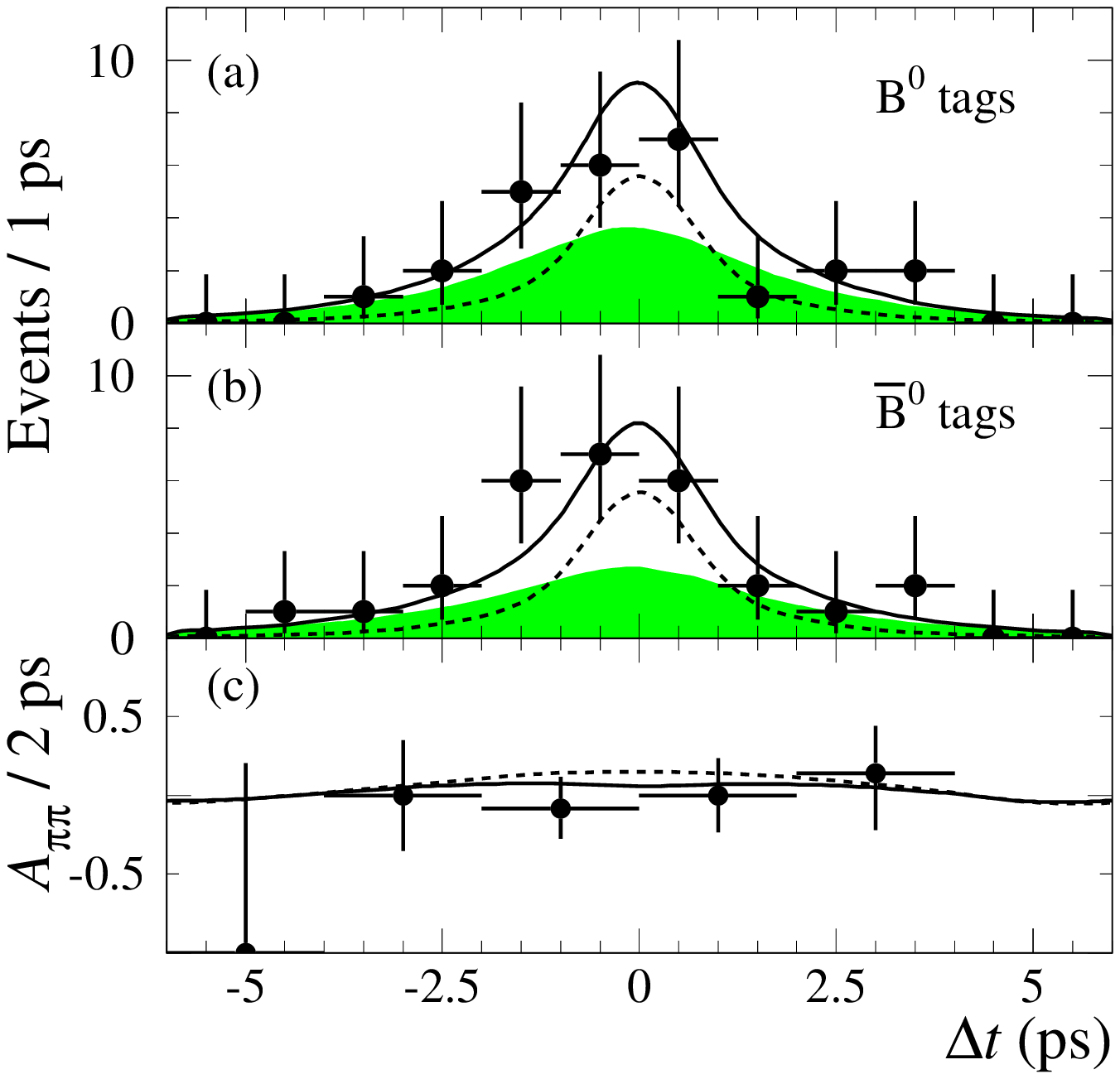}
\end{center}
\caption{%
Left: Distributions of $\mes$ and $\de$ (unshaded histograms)
for events enhanced in signal (a), (b) $\pi\pi$ and (c), (d) $K\pi$
decays. Solid curves represent projections of the maximum likelihood fit
result, while dashed curves represent $q\bar{q}$ and
$\pi\pi\leftrightarrow K\pi$ cross-feed background.  Shaded histograms
show the subset of events that are tagged. Right: Distributions of
$\deltat$ for events enhanced in signal $\pi\pi$ decays. Figures (a) and
(b) show events (points with errors) with $\Btag=\Bz$ or $\Bzb$.  Solid
curves represent projections of the likelihood fit, dashed curves
represent the sum of $q\bar{q}$ and $K\pi$ background, and the shaded
region represents the contribution from signal $\pi\pi$.  Figure (c)
shows ${\cal A}_{\pi\pi}(\deltat)$ for data (points with errors), as
well as fit projections for signal and background (solid curve), and
signal only (dashed curve).} 
\label{fig:prplots}
\end{figure}

We select $B\to \hh$ candiates  in the region $5.2 < \mes < 5.3\gevcc$
and $\left|\de\right|<0.15\gev$ and apply requirements  on track
multiplicity and event topology. The total number of events satisfying
these criteria is $9741$. This sample contains $97\%$ background, mostly
due to random combinations of tracks produced in $\epem\to q\bar{q}$
events. We extract signal and background yields for $\pip\pim$,  
$\Kp\pim$, and $\Kp\Km$ decays, and the amplitudes of the $\pi\pi$ sine
($\spipi$) and cosine ($\cpipi$) oscillation terms simultaneously from
an unbinned likelihood fit.  We parameterize the $K\pi$ component in
terms of the total yield and the $\CP$-violating charge asymmetry ${\cal
  A}_{K\pi} \equiv (N_{\Km\pip} - N_{\Kp\pim})(N_{\Km\pip} +
N_{\Kp\pim})$. Background parameters are determined from \mes and \de\
sideband regions.  

Discrimination between signal and background is based on $\mes$, $\de$,
and a Fisher discriminant ${\cal F}$~\cite{twobodyPRL} constructed from 
the scalar sum of the CM momenta of tracks and photons  (excluding
tracks from the \Brec candidate) and between pions and kaon tracks on
the Cherenkov angle measurement from the DIRC. The inclusion of events
with no flavor tag improves the signal yield estimates and provides a
larger sample for determining background shape parameters in the
likelihood fit.  

There are $18$ free parameters in the fit.  In addition to the
\CP-violating parameters $\spipi$, $\cpipi$, and ${\cal A}_{K\pi}$ (3),
the fit determines signal and background yields (6), the background
$K\pi$ charge asymmetry (1), and parameters describing the background
shapes in $\mes$, $\de$, and ${\cal F}$ (8).  We fix $\tau$ and
$\deltamd$~\cite{PDG2000}.  
The $\deltat$ PDF for signal $\pip\pim$ decays is given by
Eq.~\ref{eq:timedist}, modified to include mistags
and convolved with the signal resolution function.  The $\deltat$ PDF 
for signal $K\pi$ events takes into account $\Bz$--$\Bzb$ mixing and 
$\Bz\to\Kp\Km$ decays are parameterized as an exponential convolved with 
the resolution function. 

Figure~\ref{fig:prplots} shows distributions of $\mes$ and $\de$ for
events enhanced in signal decays based on likelihood ratios and the
$\deltat$ distributions and \CP asymmetry ${\cal A}_{\pi\pi}(\deltat) =
(N_{\Bz}(\deltat) - N_{\Bzb}(\deltat))/(N_{\Bz}(\deltat) +
N_{\Bzb}(\deltat))$ for tagged events enhanced in signal $\pi\pi$ decays 
(approximately $24$ $\pi\pi$, $22$ $q\bar{q}$, and $5$ $K\pi$ events
satisfy this selection).

In conclusion, in a sample of $33$ million $\BB$, we find
$65^{+12}_{-11}$ $\pi\pi$, $217\pm 18$ $K\pi$, and $4.3^{+6.3}_{-4.3}$
$KK$ events.  These yields are consistent with the branching fractions
reported in Ref.~\cite{twobodyPRL}. We measure the following $\CP$
parameters: 
\begin{eqnarray*}
\spipi & = & 0.03^{+0.53}_{-0.56}\mbox{ (stat)} \pm 0.11\mbox{ (syst)}
,\ \ \
{\cal A}_{K\pi}\  =\  -0.07 \pm 0.08\mbox{ (stat)} \pm 0.02\mbox{ (syst)},\\
\cpipi & = & -0.25^{+0.45}_{-0.47}\mbox{ (stat)} \pm 0.14\mbox{ (syst)}.
\end{eqnarray*}
The correlation between $\spipi$ and $\cpipi$ is $-21\%$, while ${\cal
  A}_{K\pi}$ is uncorrelated with $\spipi$ and $\cpipi$. 
Systematic errors on $\spipi$, $\cpipi$, and ${\cal A}_{K\pi}$ arise
  primarily from uncertainties in PDF shapes, tagging efficiencies and
  dilutions, $\tau$, and $\deltamd$.   

\section{Summary and Outlook}
We have observed \CP violation in the neutral $B$ system at the 4.1
$\sigma$ level using a sample of fully reconstructed $B^0$ decays to \CP
eigenstates.  With the same novel technique of time-dependent
measurements, we have determined the $B^+$ and $B^0$ lifetimes and the
\BzBzb mixing frequency \deltamd with high precision. In addition, we
have presented the first measurement of the time-dependent asymmetry in
$B^0\to\pip\pim$ decays. All results are limited by the data sample size
and we expect improved measurements from the rapidly growing \babar\ data
sample in the near future especially for the \CP violating asymmetries.  

\bibliography{hf9_prell}

\end{document}